\documentclass[acmtog,screen]{acmart}

\usepackage{booktabs} 

\usepackage{caption}
\usepackage{subcaption}
\usepackage{color}

\usepackage[ruled]{algorithm2e} 

\SetAlFnt{\small}
\SetAlCapFnt{\small}
\SetAlCapNameFnt{\small}
\SetAlCapHSkip{0pt}

\setcopyright{cc}
\setcctype{by}
\acmJournal{TOG}
\acmYear{2026} \acmVolume{45} \acmNumber{6} \acmArticle{259}
\acmMonth{12} \acmDOI{10.1145/3842571}

\begin{document}
\title{Macrofacet Theory for Gaussian Process Statistical Surfaces}

\author{Minghao Huang}
\orcid{0009-0002-7086-8190}
\email{minghao\_huang@ucsb.edu}
\affiliation{
    \institution{University of California Santa Barbara}
    \city{Santa Barbara}
    \country{United States of America}
}
\author{Yuang Cui}
\orcid{0009-0006-8983-7844}
\email{yuangcui@outlook.com}
\affiliation{
    \institution{Anhui Science and Technology University}
    \city{Anhui}
    \country{China}
}
\author{Beibei Wang}
\orcid{0000-0001-8943-8364}
\email{beibei.wang@nju.edu.cn}
\affiliation{
    \institution{Nanjing University}
    \city{Suzhou}
    \country{China}
}
\author{Lingqi Yan}
\orcid{0000-0002-9379-094X}
\email{lingqi.yan@mbzuai.ac.ae}
\affiliation{
    \institution{Mohamed bin Zayed University of Artificial Intelligence}
    \city{Abu Dhabi}
    \country{United Arab Emirates}
}

\begin{abstract}
We present macrofacet theory to extend microfacet theory from the micro-space to the macro-space. This is achieved by transforming surfaces into volumetric representations that preserve microfacet characteristics. Therefore, we formulate a macroscopic microfacet model using a classic exponential participating medium. Meanwhile, we observe that traditional microfacet models are equivalent to Gaussian processes by definition but ignore the correlation along the geometric normal of the macro-surface. We extend microfacet theory to address this limitation. Our formulation represents Gaussian process implicit surfaces in a statistical manner, which we refer to as Gaussian process statistical surfaces. As a result, our approach converts Gaussian process statistical surfaces into classic exponential media to render surfaces, volumes and in-betweens without realizations. This enables efficient rendering and improves performance compared to realization-based approaches, while theoretically bridging microfacet models and Gaussian processes. Moreover, our approach is easy to implement.
\end{abstract}

%
%


%
%


\begin{teaserfigure}
  \includegraphics[width=\textwidth]{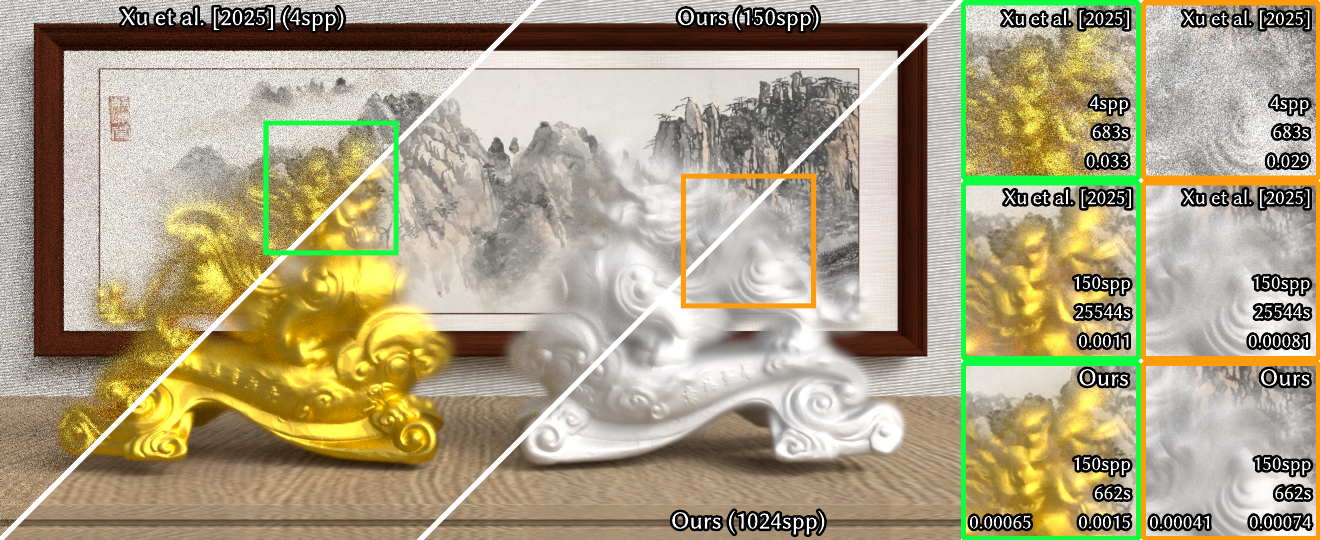}
  \caption{Our macrofacet theory connects microfacet theory and Gaussian process implicit surfaces theoretically. We use classic exponential participating media to represent surfaces, volumes and in-betweens without realizations of Gaussian processes. We compare our method with previous work by Xu et al.~\shortcite{SCNoise} at equal time and equal samples per pixel (spp). The rightmost two columns show magnified insets comparing two approaches. The spp, time for the full image and the mean squared error (MSE) for image patches are shown in corners. The reference for the MSE on the left is the converged result by ours, showing a significant improvement in rendering efficiency. The reference for the MSE on the right is the converged result by Xu et al.~\shortcite{SCNoise}, showing a good visual consistency although we make a different independence assumption, as discussed in Section~\ref{sec:discuss}.}
  \label{fig:teaser}
\end{teaserfigure}

\maketitle

\section{INTRODUCTION}

Surfaces and volumes have traditionally been treated as two distinct fields in rendering. Surface appearance is typically defined by explicit geometries, while volumetric appearance is modeled through stochastic light transport governed by absorption and scattering with particles. This separation has led to two largely independent theoretical and practical pipelines, despite the fact that many real-world appearances such as porous materials \cite{vmf} and partially coherent structures --- lie somewhere in between. Bridging the gap between surface and volumetric appearance is important both theoretically and practically. From a theoretical perspective, it raises fundamental questions about how geometries and particles should be represented across scales. From a practical standpoint, a unified treatment promises more robust and efficient rendering of materials without manually switching between surface and volumetric rendering techniques.

Microfacet theory \cite{beckmann1963scattering, CookTorrance} is widely used to represent surfaces, while microflake theory \cite{microflake} is widely used to represent participating media. Dupuy et al.~\shortcite{Dupuy} unify the two theories by interpreting a microfacet surface as an anisotropic participating medium. Their unification, however, stays within the micro-scale and provides no continuous transition from a surface to a volume. It also inherits the height field assumption of microfacet theory, and therefore cannot represent micro-geometries with holes and overlaps, such as porous materials.

Seyb et al.~\shortcite{GPIS} and Xu et al.~\shortcite{SCNoise} use Gaussian process implicit surfaces (GPISes) to unify surface, volume and in-between representations. However, rendering GPISes requires realizations of Gaussian processes, which are computationally expensive and difficult to implement. In fact, we observe that the traditional microfacet model is defined as a 2D Gaussian process on the $xy$-plane which is the macro-surface \cite{Beckmann}. Therefore, it is a 2D GPIS without considering the correlation along the geometric normal of the macro-surface, or the $z$-axis. The microfacet model uses the height distribution, the normal distribution function and the shadow masking to describe the 2D GPIS statistically to avoid realizations. Inspired by this, we adopt a statistical perspective for modeling 3D GPISes, which we refer to as Gaussian process statistical surfaces (GPSSes). Hence, we can render GPSSes without massive realizations.

\begin{figure}[t]
    \centering
    \includegraphics[width=\linewidth]{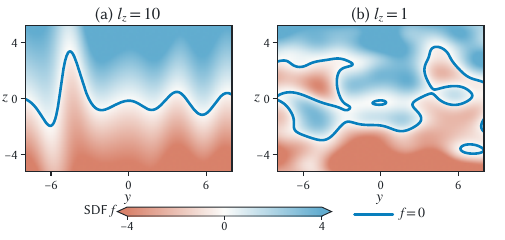}
    \caption{GPSSes with different correlations along the $z$-axis. These two figures show signed distance fields (SDFs) in $yz$-slices of realizations of (a) a strongly anisotropic GPSS with $l_z=10$ and (b) an isotropic one with $l_z=1$. The blue lines are the zero level sets, which are indeed the realizations of GPSSes. The strongly anisotropic GPSS (a) is a height field, while the isotropic one (b) is no longer a height field, having normals pointing downward.}
    \label{fig:diff_lz}
\end{figure}

To model 3D GPSSes, we use classic exponential participating media to extend microfacet models from the micro-space to the macro-space. This perspective also explains, in terms of Gaussian processes, why microflake-based interpretations of microfacet theory \cite{Dupuy, 16Heitz} are restricted to height fields: their essences are still 2D GPSSes. Once moved into 3D, a 2D GPSS becomes a fully anisotropic 3D GPSS whose correlation along the geometric normal, or the $z$-axis, is infinite, as shown in Figure~\ref{fig:diff_lz}. We instead extend microfacet theory to the macro-space and support general 3D GPSSes, where the correlations on all axes are arbitrary.

Based on the foregoing analysis, we adopt a statistical perspective for modeling GPISes and propose macrofacet theory to: 1) represent GPSSes in classic exponential participating media; 2) extend microfacet theory to describe general 3D GPSSes. Therefore, we unify surface, volume and in-between representations using macrofacets. We extend a macro-surface to a volume in order to transform microfacet theory to microflake theory in the macro-space. We connect our height field macrofacet to the fully anisotropic GPSS. Also, in order to map our generalized macrofacet to the general GPSS, we analytically compute the transmittance, normal distribution function and phase function of the corresponding participating media within a de-correlation assumption that the signed distance field and gradient at the next intersection are independent of those at the last intersection. As a consequence, we can render GPSSes by any classic exponential participating media rendering techniques \emph{without realization}. Hence, our approach is computationally efficient and easy to implement. Code for this paper is at \url{https://github.com/BakaSea/Macrofacet}.

\section{RELATED WORK}

\subsection{Microfacet Surface}

Beckmann and Spizzichino~\shortcite{beckmann1963scattering} propose the microfacet model for reflection on rough surfaces. They average over reflectance from a statistical representation of the micro-geometry of the surface consisting of specular microfacets. This model is introduced to the graphics community by Cook and Torrance~\shortcite{CookTorrance} and then extended to rough dielectrics \cite{Stam}. Walter et al.~\shortcite{GGX} use the Smith's assumption \cite{Smith} to apply more normal distribution functions to the model, such as the GGX distribution.

Different normal distribution functions result in different appearances of the surface. In the beginning, the Beckmann distribution \cite{Beckmann, CookTorrance} was widely used. Trowbridge and Reitz~\shortcite{Trowbridge:75} propose a distribution on half-ellipsoids. Walter et al.~\shortcite{GGX} adopt it as the GGX distribution to achieve a more realistic appearance.

The typical microfacet model only calculates the reflectance from a single bounce of a light ray intersecting the micro-geometry of the surface, which results in an energy loss, especially when the roughness is high. Heitz et al.~\shortcite{16Heitz} treat microfacets as randomly distributed microflakes using a height distribution so that they can trace multiple bounce rays in a random walk solution. Wang et al.~\shortcite{Position-free} assume that microflakes are position-free to avoid noise from the height distribution. Cui et al.~\shortcite{Invariance-principle} further introduce the invariance principle to generalize the shadow masking function from a single bounce to an entire path to reduce noise. Since we use a volume to represent materials, our macrofacet supports multiple bounces naturally.

\subsection{Gaussian Processes}

Gaussian processes are used to model distributions over functions. They are widely used in machine learning \cite{GPML}, signal processing \cite{GPSP} and other areas. In computer graphics, they have been applied for surface reconstruction \cite{williams2007gaussian, MartenGaussian}, where a Gaussian process is conditioned on a set of point observations and the surface is extracted from the zero-crossing of the Gaussian process mean.

Seyb et al.~\shortcite{GPIS} propose a light transport framework using Gaussian processes to represent stochastic geometries. They realize Gaussian processes to obtain implicit surfaces on-the-fly and ensemble average over the light transport on these realizations. This unifies the representations of surfaces, volumes and even the in-betweens, which show the macro-scale uncertainty. However, because of the heavy computation of realizations, it requires a significant amount of time to render results in spite of their Renewal and Renewal+ models. Xu et al.~\shortcite{SCNoise} introduce sparse convolution noise to approximate Gaussian processes to reduce the computation of realizations. They also utilize next-event estimation to further reduce noise. Shi et al.~\shortcite{Shi} enable next-event estimation on heightfield Gaussian process implicit surfaces. Nevertheless, both of them require a large number of realizations of Gaussian processes while ray marching, resulting in low efficiency. Chen et al.~\shortcite{BSBI} introduce stratified Bernoulli impulses on the basis of sparse convolution noise to accelerate ray-surface intersections, but they are still realization-based methods. Moreover, these methods are hard to implement in current rendering engines. Our macrofacet is a classic participating medium, so it has a high rendering efficiency and is easily implemented in current rendering engines.

\subsection{Participating Media}

The standard form of the radiative transport equation \cite{VRTE} is usually stated for spherical or randomly oriented particles, which is not the case for anisotropic media. Jakob et al.~\shortcite{microflake} propose a physically-based radiative transfer framework called microflake for anisotropic media \cite{kuscer, Williams_1978}. The volume scattering model within this framework is analogous to the microfacet model, using oriented non-spherical particles. It is widely used in woven \cite{Zhao11, Zhao12} and cloth \cite{MicroflakeCloth} materials. Heitz et al.~\shortcite{SGGX} introduce the SGGX distribution to represent spatially-varying properties of anisotropic media based on microflake theory. Dupuy et al.~\shortcite{Dupuy} assume a semi-infinite homogeneous exponential-free-path medium to unify microfacet and microflake theories. However, their work is still an extension of traditional microfacet theory, where microfacet models are fully anisotropic 3D GPSSes. Therefore, it cannot handle general 3D GPSSes. Our macrofacet theory extends microfacet theory to support 3D GPSSes.

Miller et al.~\shortcite{Miller:VOS:2024} develop a theory for the stochastic representation of opaque solids as classic exponential participating media. Nevertheless, their work is based on a continuous-time discrete-space Markov process, which is an uncorrelated stochastic process and not the case for a Gaussian process with a squared exponential kernel \cite{GPIS}. Furthermore, their projected area relies on a prescribed distribution of normals. Our macrofacet theory does not have this restriction and derives the projected area from the stochastic geometry.

\section{BACKGROUND AND OVERVIEW}

In this section, we introduce the background knowledge of microfacet and microflake theories, Gaussian process implicit surfaces, and give an overview of our proposed macrofacet theory. The notations that will be used throughout our paper are listed in Table~\ref{tab:notations}.

\begin{table}[t]
    \centering
    \caption{Notations.}
    \begin{tabular}{|c|c|}
        \hline
        \multicolumn{2}{|c|}{Mathematical notation}\\
        \hline
        $\Omega$ & full spherical domain \\
        $\mu$ & mean \\
        $\sigma^2$ & variance \\
        $l$ & correlation \\
        $\kappa$ & covariance kernel \\
        $\phi(x; \mu, \sigma^2)$ & Gaussian probability density function \\
        & with mean $\mu$ and variance $\sigma^2$ \\
        $\Phi(x; \mu, \sigma^2)$ & Gaussian cumulative density function \\
        & with mean $\mu$ and variance $\sigma^2$ \\
        $\omega_1\cdot\omega_2$ & dot product \\
        $|\omega_1\cdot\omega_2|$ & absolute value of the dot product \\
        $\langle\omega_1,\omega_2\rangle$ & clamped dot product \\
        \hline
        \multicolumn{2}{|c|}{Physical quantities}\\
        \hline
        $f$ & signed distance field \\
        $g$ & gradient of a signed distance field \\
        $\alpha$ & roughness \\
        $\omega_g=(0, 0, 1)$ & geometric normal \\
        $\omega_m$ & microfacet normal \\
        $\omega_i$ & incident direction \\
        $\omega_o$ & outgoing direction \\
        $h$ & microsurface height on the $z$-axis \\
        $P^1(h)$ & height distribution \\
        $C^1(h)$ & cumulative height distribution \\
        $\Lambda(\omega)$ & the Smith Lambda function \\
        $D(\omega_m)$ & normal distribution function \\
        $F(\omega_o, \omega_m)$ & Fresnel term \\
        $D_{\omega_o}(\omega_m)$ & visible normals' distribution \\
        $\rho(h)$ & Microflake density \\
        $\sigma(\omega_o)$ & Microflake projected area \\
        $\sigma_t(\omega_o, h)$ & Microflake extinction coefficient \\
        $p(\omega_o, \omega_i)$ & Microflake phase function \\
        \hline
    \end{tabular}
    \label{tab:notations}
\end{table}

\subsection{Microfacet and Microflake Theories}

Microfacet theory describes a statistical model at the micro-scale with a height distribution $P^1(h)$ and a normal distribution function $D(\omega_m)$ of micro-surfaces. The Smith's assumption \cite{Smith} de-correlates $P^1(h)$ and $D(\omega_m)$, so we can choose them independently. The height distribution can be chosen arbitrarily \cite{16Heitz, Position-free, Invariance-principle, 22Bitterli}. However, the SDF distribution of any point in the space is a Gaussian distribution if it is generated by a Gaussian process. Therefore, in order to match our method to GPISes, we choose the Gaussian distribution with variance $\sigma^2$ for $P^1(h)$:
\begin{equation}
\label{eqn:height}
P^1(h)=\phi(h;0,\sigma^2)=\frac{1}{\sqrt{2\pi}\sigma}e^{-\frac{h^2}{2\sigma^2}},
\end{equation}
where $h\in(-\infty,+\infty)$ and $h=0$ means the average plane of the surface.
There are many types of normal distribution functions $D(\omega_m)$, usually referred to as NDFs. Two mainly used distributions are the Beckmann and GGX distributions \cite{GGX}. They are defined by the roughness $\alpha$ equivalent to Equation~\eqref{eqn:roughness}.

Dupuy et al.~\shortcite{Dupuy} and Heitz et al.~\shortcite{16Heitz} use microflake theory \cite{microflake,SGGX} to describe microfacet theory. The light transport in an anisotropic participating medium is described by the anisotropic radiative transfer equation \cite{microflake}:
\begin{equation}
(\omega_o\cdot\nabla)L(\omega_o)+\sigma_t(\omega_o)L(\omega_o)=\sigma_s(\omega_o)\int_{\Omega}p(\omega_o, \omega_i)\mathrm{d}\omega_i+Q(\omega_o).
\end{equation}
Here, $\sigma_t(\omega_o)$ is the extinction coefficient, $\sigma_s(\omega_o)$ is the scattering coefficient, $p(\omega_o, \omega_i)$ is the phase function, and $Q(\omega_o)$ is the radiance emitted by the medium. We let $\sigma_s(\omega_o)=\sigma_t(\omega_o)$ to incorporate albedo in the phase function. Hence, we can define an anisotropic participating medium by determining the extinction coefficient and phase function. The microflake model parameterizes them using a distribution of microflakes, which consists of density and NDF.

\begin{figure}[t]
    \centering
    \includegraphics[width=0.9\linewidth]{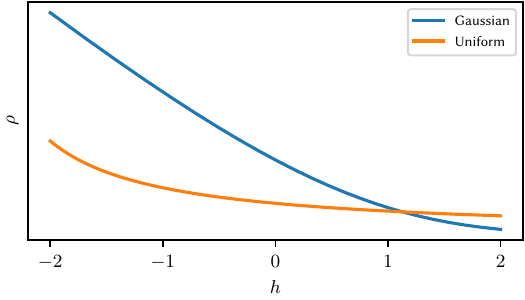}
    \caption{The density decreases monotonically with increasing height $h$ for Gaussian distributions and uniform distributions. The blue line is a Gaussian distribution with the mean $\mu=0$ and the variance $\sigma^2=1$. The orange line is a uniform distribution within $[-3, 3]$.}
    \label{fig:monotonic_density}
\end{figure}

The extinction coefficient consists of density and projected area. The density of the volume associated with a Smith microfacet surface is defined as
\begin{equation}
\label{eqn:h_density}
\rho(h)=\frac{P^1(h)}{C^1(h)},
\end{equation}
where $C^1(h)$ is the cumulative density function of $P^1(h)$.
It is monotonic rather than symmetric for Gaussian distributions and uniform distributions, as shown in Figure~\ref{fig:monotonic_density}.
The corresponding projected area is
\begin{equation}
\sigma(\omega_o)=\int_{\Omega}\langle -\omega_o, \omega_m\rangle D(\omega_m)\mathrm{d}\omega_m=\Lambda(\omega_o)\cos\theta_o,
\end{equation}
where $\cos\theta_o$ is the cosine of the angle between the outgoing direction $\omega_o$ and the geometric normal $\omega_g$. We clarify that the outgoing direction $\omega_o$ is always pointing to the origin of the incoming direction $\omega_i$. The extinction coefficient is the product of the density and projected area: 
\begin{equation}
\label{eqn:micro_sigma_t}
\sigma_t(\omega_o, h)=\rho(h)\sigma(\omega_o)=\frac{P^1(h)}{C^1(h)}\Lambda(\omega_o)\cos\theta_o.
\end{equation}
Given the extinction coefficient, the transmittance can be computed as the remaining energy when a ray travels at distance $t$ in the medium:
\begin{equation}
\mathrm{Tr}(\omega_o, t)=\exp\left(-\int_0^t\sigma_t(\omega_o, h(t))\mathrm{d}t\right).
\end{equation}
This equation establishes the equivalence between the extinction coefficient and transmittance, indicating that either quantity can be derived from the other.

The phase function of a volume, describing how a ray is scattered in the medium, depends on its material. In this paper, we focus on the conductor material, for which we can obtain a conductor phase function: 
\begin{equation}
\label{eqn:phase}
p(\omega_o, \omega_i)=\frac{F(-\omega_o, \omega_m)D_{\omega_o}(\omega_m)}{4|-\omega_o\cdot\omega_m |},
\end{equation}
where $\omega_m=\frac{-\omega_o+\omega_i}{||-\omega_o+\omega_i||}$ is the half vector. The visible normals' distribution (vNDF) $D_{\omega_o}(\omega_m)$ is defined as
\begin{equation}
\label{eqn:microfacet_vndf}
D_{\omega_o}(\omega_m)=\frac{\langle-\omega_o,\omega_m\rangle D(\omega_m)}{\int_{\Omega}\langle -\omega_o, \omega_m\rangle D(\omega_m)\mathrm{d}\omega_m}=\frac{\langle-\omega_o,\omega_m\rangle D(\omega_m)}{\Lambda(\omega_o)\cos\omega_o}.
\end{equation}
This phase function satisfies both reciprocity and energy conservation. According to Equation~\eqref{eqn:phase} and Equation~\eqref{eqn:microfacet_vndf}, we know the relation between the NDF and phase function. Therefore, when we derive the phase function for the macrofacet later, we need to know the NDF.

\subsection{Gaussian Process Implicit Surfaces}

Microfacet and microflake theories are based on the height field assumption of the micro-geometry, and therefore are unable to handle the correlation along the geometric normal of the macro-surface. Gaussian process implicit surfaces generalize them. A Gaussian process $f(\mathbf{x})\sim \mathcal{GP}(\mu(\mathbf{x}), \kappa(\mathbf{x}, \mathbf{y}))_{\mathbb{R}^3}$ is a distribution over functions $f$ characterized by the mean function $\mu(\mathbf{x})=\mathbb{E}(f(\mathbf{x}))$ and the covariance kernel $\kappa(\mathbf{x}, \mathbf{y})=\mathrm{Cov}(f(\mathbf{x}), f(\mathbf{y}))$. There are many types of covariance kernels. We focus on the squared exponential (SE) kernel because it is widely used and previous works \cite{GPIS, SCNoise} use it to represent volume-type and in-between Gaussian process implicit surfaces. It is a stationary kernel and defined as
\begin{equation}
\label{eqn:se_kernel}
\kappa(\mathbf{x}, \mathbf{y})=\sigma^2\exp\left(-\frac{1}{2}(\mathbf{x}-\mathbf{y})^T\mathrm{diag}(l_x^2, l_y^2, l_z^2)^{-1}(\mathbf{x}-\mathbf{y})\right),
\end{equation}
where $\sigma^2$ is the variance. $l_x$, $l_y$ and $l_z$ are the correlation along the $x$-, $y$- and $z$-axes respectively. When $l_x=l_y=l_z$, the covariance kernel is isotropic. Otherwise, it is anisotropic. The variance and correlation are independent variables.

Due to the linearity of the derivative operator, the gradient $g(\mathbf{x})$ of a Gaussian process $f(\mathbf{x})$ is again a Gaussian process. The joint value-derivative distribution is
\begin{equation}
\label{eqn:joint_fg}
\begin{pmatrix}
f(\mathbf{x})\\
g(\mathbf{x})
\end{pmatrix}
\sim\mathcal{N}\left(
\begin{pmatrix}
\mu(\mathbf{x})\\
\nabla\mu(\mathbf{x})
\end{pmatrix},
\begin{pmatrix}
\kappa(\mathbf{x}, \mathbf{y}) & \nabla_\mathbf{y}^\top\kappa(\mathbf{x}, \mathbf{y})\\
\nabla_\mathbf{x}\kappa(\mathbf{x},\mathbf{y}) & \nabla_\mathbf{x}\nabla_\mathbf{y}^\top\kappa(\mathbf{x}, \mathbf{y})
\end{pmatrix}
\right),
\end{equation}
where $\nabla_\mathbf{x}\coloneq (\partial/\partial\mathbf{x}_x, \partial/\partial\mathbf{x}_y, \partial/\partial\mathbf{x}_z)^\top$ denotes the gradient operator with respect to $\mathbf{x}$. Note that when $\mathbf{x}$ is a single random variable, $f(\mathbf{x})$ and $g(\mathbf{x})$ are naturally independent in the computation.

\begin{figure*}[t]
    \centering
    \includegraphics[width=\linewidth]{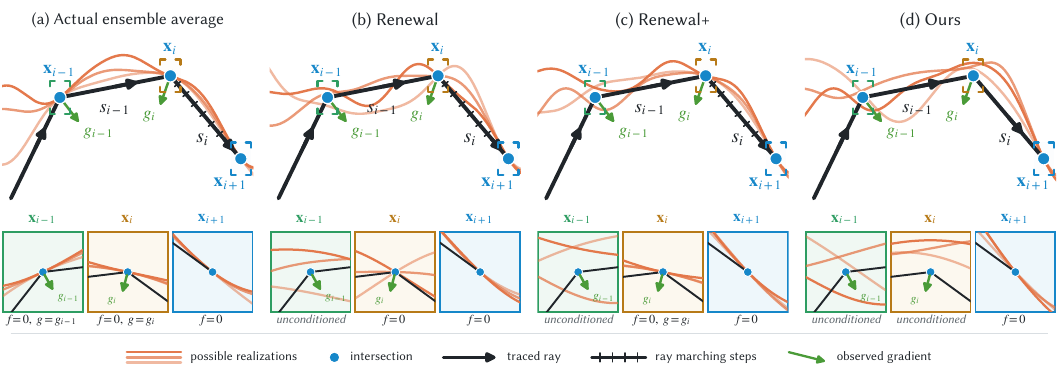}
    \caption{Different independence assumptions between the macrofacet and GPIS approaches. A sub-path consists of 2 segments $s_{i-1}$, $s_i$, and 3 intersections $\mathbf{x}_{i-1}$, $\mathbf{x}_i$, $\mathbf{x}_{i+1}$. The key differences are shown in the insets. When marching the last segment $s_i$ to find the next intersection $\mathbf{x}_{i+1}$, (a) the actual ensemble average guarantees that all possible realizations cross through all intersections, match the gradient at every intersection and do not occlude any segments. (b) The Renewal model only conditions on the previous intersection $\mathbf{x}_i$ and ignores earlier segments and intersections. It guarantees that all possible realizations cross through $\mathbf{x}_i$ and $\mathbf{x}_{i+1}$, and do not occlude $s_i$. (c) The Renewal+ model conditions on the gradient $g_i$ at $\mathbf{x}_i$ in addition to the Renewal model. It guarantees that all possible realizations cross through $\mathbf{x}_i$ and $\mathbf{x}_{i+1}$, match $g_i$ at $\mathbf{x}_i$, and do not occlude $s_i$. (d) The macrofacet is independent of all other segments and intersections except for $\mathbf{x}_{i+1}$. It guarantees that all possible realizations cross through $\mathbf{x}_{i+1}$, and do not occlude $s_i$. The impact of different independence assumptions is shown in Figure~\ref{fig:impact_independent}.}
    \label{fig:indpendent_assumption}
\end{figure*}

A Gaussian process can be conditioned on a set of observations $(\mathcal{X}, m)$, where $\mathcal{X}$ is a set of points on the input domain and $m$ are the observed values at those locations. After observing $(\mathcal{X}, m)$, the conditioned posterior process is again a Gaussian, with mean and covariance:
\begin{equation}
\begin{split}
\label{eqn:condition_gp}
\mu_{\mid \mathcal{X}}(\mathbf{x})=&\mu(\mathbf{x})+\kappa(\mathbf{x}, \mathcal{X})\kappa(\mathcal{X}, \mathcal{X})^{-1}(m-\mu(\mathcal{X})),\\
\kappa_{\mid \mathcal{X}}(\mathbf{x}, \mathbf{y})=&\kappa(\mathbf{x}, \mathbf{x})-\kappa(\mathbf{x}, \mathcal{X})\kappa(\mathcal{X}, \mathcal{X})^{-1}\kappa(\mathcal{X},\mathbf{x}).
\end{split}
\end{equation}

A Gaussian Process Implicit Surface (GPIS) is the zero level set of a Gaussian process. This implicit surface is stochastic: every realization $f$ drawn from the Gaussian process generates a different surface where $f(\mathbf{x})=0$. To render a GPIS, previous works \cite{GPIS, SCNoise} perform realizations on-the-fly for every path. They use ray marching to find zero-crossings of realizations. The total radiance received is the ensemble average over all possible realizations.

However, realizations of a GPIS are extremely computationally expensive. It takes $\mathcal{O}(m^3)$ to compute a realization for a path, where $m$ is the number of steps taken along the path. The problem of the complexity is that the realization needs to be consistent across bounces of a path. To address this problem, Seyb et al.~\shortcite{GPIS} propose Renewal and Renewal+ models. These models only maintain such consistency to the previous bounce, as shown in Figure~\ref{fig:indpendent_assumption}. Therefore, they reduce the time complexity to $\mathcal{O}(n^3)$, where $n$ is the number of steps taken along the current segment. Xu et al.~\shortcite{SCNoise} use sparse convolution noise to approximate Gaussian processes so that they can evaluate the value of each point in $\mathcal{O}(1)$. Nevertheless, they use brute force ray marching to find zero-crossings, resulting in the need of $\mathcal{O}(n)$ to find intersections. It will require a significant amount of time to find intersections if the segment is long. Notably, all of them ignore the correlation between the current segment and the previous segments except for the last intersection, so they are not the actual ensemble average of the GPIS.

A GPIS exhibits three types of appearance: surface, volume and in-between, determined by its mean function and covariance kernel \cite{GPIS}. In-betweens present macro-scale uncertainty as ``fuzziness''. When the mean function is the signed distance field (SDF), the GPIS appears like a surface or an in-between. When the mean function is a constant, it appears like a homogeneous medium. The covariance kernel affects not only the roughness but also the fuzziness of the appearance. The roughness can be derived from the second derivative of the covariance kernel \cite{PBRTV3}. Here is for the SE kernel:
\begin{equation}
\label{eqn:roughness}
\alpha=\sqrt{-2\kappa''(0)}=\sqrt{2}\frac{\sigma}{l}.
\end{equation}

\subsection{Overview}

There are two main reasons why previous works \cite{GPIS, SCNoise} are slow: 1) they require expensive computations to find ray intersections; 2) they perform individual realizations for every path, leading to a low convergence efficiency.

Microfacet models describe rough surfaces through a statistical representation of the micro-geometry, while volumetric models likewise rely on a statistical formulation to avoid explicit simulation of individual particle interactions. This shared statistical perspective suggests a similar treatment for the GPIS. By formulating the GPIS in a statistical manner, rather than relying on explicit realizations, rendering efficiency can be significantly improved. We refer to this representation as the Gaussian process statistical surface (GPSS).

We observe that the Beckmann microfacet model can be interpreted as a 2D GPSS. However, it only describes the fully anisotropic GPSS rather than the general one when it is moved to 3D. Building on this observation, we propose macrofacet theory, which extends microfacet models from the microscopic scale to the macroscopic scale and supports general 3D GPSSes. To describe microfacet models in the macroscopic scale, we use classic exponential participating media. It is modeled by the extinction coefficient and phase function. We derive analytical solutions for them from statistical analysis for general GPSSes, which is far from trivial. As a consequence, GPSSes can be rendered using volumetric rendering techniques, achieving significantly higher efficiency than realization-based approaches.

\section{MACROFACET}

In this section, we propose our macrofacet theory and show how it can connect microfacet theory and the GPSS.

\begin{figure}[t]
    \centering
    \includegraphics[width=\linewidth]{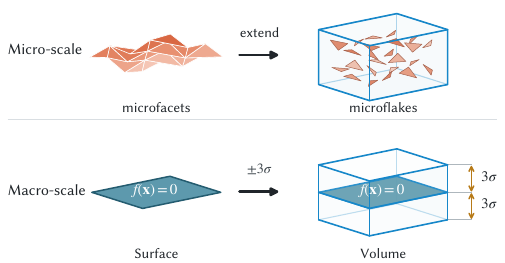}
    \caption{Extension of a macro-surface. The top row illustrates the micro-scale extension of the macrofacet, while the bottom row illustrates the macro-scale extension of the macrofacet. Originally, the object is a surface at the macro-scale and microfacet at the micro-scale. We stretch the macro-surface upwards and downwards by the length of $3\sigma$. Consequently, the macro-surface becomes a shell volume. At the same time, microfacets become microflakes floating inside the shell.}
    \label{fig:extension}
\end{figure}

The core idea of macrofacet theory is to first take micro-surfaces from the micro-space to the macro-space. Meanwhile, the total number of micro-surfaces is maintained. Considering a volume containing many particles, how we stretch this volume does not change the total number of particles. In order to achieve this, we extend the original macro-surface in the $z$-axis, the geometric normal, to get a shell. The distance from the upper/lower face to the original macro-surface where $f(\mathbf{x})=0$ is $3\sigma$, where $\sigma^2$ is the variance of height distribution of micro-surfaces described in Equation~\eqref{eqn:height}. Since the probability outside $[-3\sigma, 3\sigma]$ of a Gaussian distribution is low enough to be ignored, it is reasonable to assume that the shell contains all micro-surfaces. The space inside the shell is a microflake volume. Then the micro-surfaces in the micro-space become microflakes in the macro-space. Figure~\ref{fig:extension} shows the extension.

Then, the height $h$ in the micro-space is mapped to the SDF $f$ in the macro-space \cite{GPIS}. Therefore, we perform parameter conversion in Equation~\eqref{eqn:h_density} and Equation~\eqref{eqn:micro_sigma_t}:
\begin{equation}
\rho(f)=\frac{P^1(f)}{C^1(f)}=\frac{\phi(f;0,\sigma^2)}{\Phi(f;0,\sigma^2)},
\end{equation}
\begin{equation}
\label{eqn:macro_sigma_t}
\sigma_t(\omega_o, f)=\frac{P^1(f)}{C^1(f)}\Lambda(\omega_o)\cos\theta_o.
\end{equation}
The normal in the macro-space is the normalized gradient of the SDF. The phase function follows Equation~\eqref{eqn:phase}. Hence, when $\sigma\to 0$, the shell degenerates to a surface and appears the same as a microfacet surface. However, the derived extinction coefficient and phase function correspond only to the height field macrofacet. In Section~\ref{sec:heightfield_macrofacet}, we will establish the connection between the height field macrofacet and fully anisotropic GPSS. In Section~\ref{sec:generalized_macrofacet}, we further introduce the generalized macrofacet to connect with the general GPSS. Interestingly, the extinction coefficient and phase function of the generalized macrofacet share the same form as those of the height field macrofacet.

\subsection{Height Field Macrofacet}
\label{sec:heightfield_macrofacet}

\begin{figure}
    \centering
    \includegraphics[width=\linewidth]{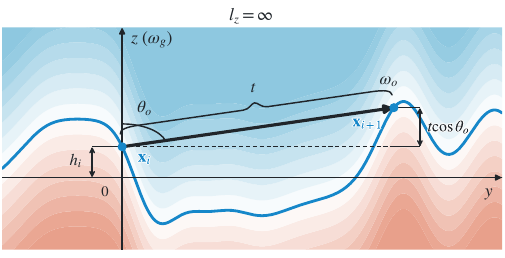}
    \caption{A cross section of a fully anisotropic GPSS intercepted by the $yz$-plane. The $z$-axis is the geometric normal $\omega_g$. The blue line is a realization of the GPSS.}
    \label{fig:heightfield_dv}
\end{figure}

We consider an infinitesimal volume element $\mathrm{d}V$, whose $z$-axis is the geometric normal, as shown in Figure~\ref{fig:heightfield_dv}. If we choose the Beckmann distribution as the macrofacet NDF, it will be a fully anisotropic GPSS where the correlation along the $z$-axis $l_z$ is infinite. In fact, the Beckmann distribution is a 2D Gaussian process with an SE kernel on the $xy$-plane \cite{Beckmann}. The roughness on the $x$-axis $\alpha_x$ and $y$-axis $\alpha_y$ is defined as $\sqrt{2}\sigma/l_x$ and $\sqrt{2}\sigma/l_y$, respectively, which is the same as Equation~\eqref{eqn:roughness}. When we put this 2D Gaussian process into 3D, the correlation along the $z$-axis will be infinity because there is only one zero-crossing along the same $z$-axis. It means that the GPSS is a height field and consistent with the microfacet assumption.

\begin{figure}[t]
    \centering
    \includegraphics[width=0.9\linewidth]{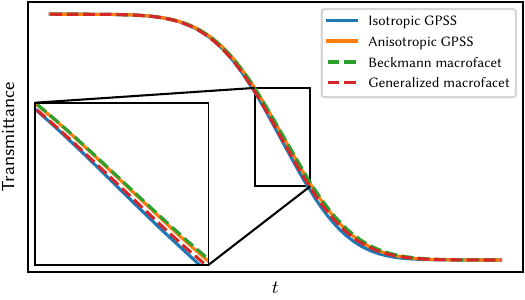}
    \caption{Comparison of transmittance between macrofacets and GPSSes. The figure shows the transmittance of a ray intersecting macrofacet and GPSS planes at a $45^\circ$ incident angle, where $-\omega_o\cdot\omega_g=\sqrt{2}/2$. The transmittance of the Beckmann macrofacet (green dashed line) matches that of the fully anisotropic GPSS (orange line) closely. The transmittance of generalized macrofacet (red dashed line) matches the one of isotropic GPSS (blue line) well in the first half, but has a slight difference in the latter half, likely attributable to the de-correlated assumption.}
    \label{fig:tr}
\end{figure}

In order to render a fully anisotropic GPSS in a classic exponential participating medium, we need to consider its phase function and transmittance. Seyb et al.~\shortcite{GPIS} show that the NDF of a fully anisotropic GPSS is the same as the Beckmann distribution, so we can use the Beckmann distribution to calculate the phase function. And calculating the transmittance of a GPSS is indeed calculating the shadow masking term $G_1(\omega_o)$. Smith~\shortcite{Smith} uses $S_\mathrm{Smith}(h_i, p_i, q_i, \omega_o)$, or $S_\mathrm{Smith}(\mathbf{x}_i, \omega_o)$ for short, to denote the probability that a point $\mathrm{x}_i$ on a fully anisotropic GPSS, of given height $h$ above the average plane, and with local slopes $p_i$, $q_i$, will not lie in shadow when the surface is illuminated with an incident light ray $\omega_o$, as shown in Figure~\ref{fig:heightfield_dv}. It can be written as the limit
\begin{equation}
S_\mathrm{Smith}(\mathrm{x}_i, \omega_o)=\lim_{t\to\infty}\mathrm{Tr}_\mathrm{Smith}(\mathrm{x}_i, \omega_o, t),
\end{equation}
where $\mathrm{Tr}_\mathrm{Smith}(\mathrm{x}_i, \omega_o, t)$ is the probability that no part of the surface between $\mathrm{x}_i$ and $\mathrm{x}_i+t\omega_o$ will intersect the ray $\omega_o$. The meaning of $\mathrm{Tr}_\mathrm{Smith}(\mathrm{x}_i, \omega_o, t)$ is equivalent to the transmittance of the ray $\omega_o$ originating from $\mathrm{x}_i$ and advancing $t$. It can be derived to
\begin{equation}
\mathrm{Tr}_\mathrm{Smith}(\mathrm{x}_i, \omega_o, t)=\exp\left(-\int_0^t\sigma_t(\omega_o, t)\mathrm{d}t\right),
\end{equation}
where $\sigma_t(\omega_o, t)$ is exactly the extinction coefficient of the volume. Smith~\shortcite{Smith} neglects the correlation between the height and slopes at $\mathrm{x}_i$ and those at $\mathrm{x}_i+t\omega_o$ in \vspace{-1mm}order to simply calculate $\sigma_t(\omega_o, t)$:
\begin{equation}
\sigma_t(\omega_o, t)=\frac{\phi(h_i+t\cos\theta_o;0,\sigma^2)}{\Phi(h_i+t\cos\theta_o;0,\sigma^2)}\Lambda(\omega_o)\cos\theta_o,
\end{equation}
where $h_i+t\cos\theta_o$ is in fact the height $h_{i+1}$ of the next intersection. It is the same as Equation~\eqref{eqn:micro_sigma_t} in the micro-space and Equation~\eqref{eqn:macro_sigma_t} in the macro-space. Therefore, the extinction coefficient in Equation~\eqref{eqn:macro_sigma_t} is consistent with the GPSS. We illustrate the comparison about the transmittance between the Beckmann macrofacet, a fully anisotropic GPSS and an isotropic GPSS in Figure~\ref{fig:tr}. It shows that the transmittance of Beckmann macrofacet matches the one of the fully anisotropic GPSS. In summary, due to the consistency with the phase function and transmittance, the Beckmann macrofacet is the same as a fully anisotropic GPSS.

Besides, with the Smith's assumption \cite{Smith}, we can choose an arbitrary NDF ignoring the density distribution of microflakes. This, instead of choosing different covariance kernels of Gaussian processes, makes macrofacet more flexible and easier for artists to understand the appearance of the macrofacet. For example, we can choose the GGX distribution as the macrofacet NDF to get a more realistic appearance than the Beckmann distribution when the shell degenerates to a surface.

\subsection{Generalized Macrofacet}
\label{sec:generalized_macrofacet}

\begin{figure}[t]
    \centering
    \includegraphics[width=\linewidth]{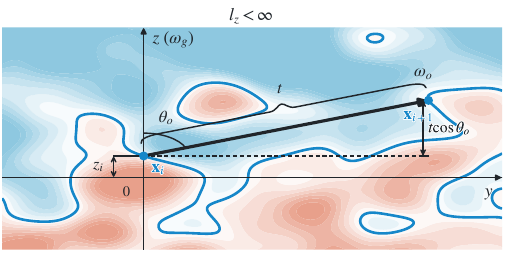}
    \caption{A cross section of a general GPSS intercepted by the $yz$-plane. The $z$-axis is the geometric normal $\omega_g$. The blue lines are the zero level sets.}
    \label{fig:generalized_dv}
\end{figure}

When the correlation along the $z$-axis is finite, the GPSS is no longer a height field, as shown in Figure~\ref{fig:diff_lz}. There are multiple zero-crossings along the same $z$-axis. And there are micro-surface normals pointing downwards, forming a microfacet surface with holes and overlaps, which is difficult to handle. Traditional microfacet theory \cite{16Heitz, Dupuy} does not handle this situation and cannot connect to a general GPSS because it only assumes a height field. In this case, we still focus on two important properties of a classic exponential participating media --- transmittance, or extinction coefficient, and phase function. We still consider an infinitesimal volume element $\mathrm{d}V$, whose $z$-axis is the geometric normal, as shown in Figure~\ref{fig:generalized_dv}.

\subsubsection{Extinction coefficient}
In order to deal with these problems, we first extend our transmittance calculation from 2D to 3D. Suppose $\mathbf{x}$ is a point in the macro-space. $f(\mathbf{x})$ is the SDF and $f(\mathbf{x})\sim \mathcal{GP}(\mu(\mathbf{x}), \kappa(\mathbf{x}, \mathbf{y}))_{\mathbb{R}^3}$, where $\mu(\mathbf{x})=z$ and $\kappa(\mathbf{x}, \mathbf{y})$ is Equation~\eqref{eqn:se_kernel}. $g(\mathbf{x})=\nabla f(\mathbf{x})$ is the gradient. Let $\mathbf{x}_i=(0, 0, z_i)$ be the initial point. $\omega_o=(\sin\theta_o\cos\phi_o, \sin\theta_o\sin\phi_o, \cos\theta_o)$ is the unit directional vector. $t$ is the travel distance. We denote the probability that a point $\mathbf{x}_i$ on a GPSS, of given SDF $f(\mathbf{x}_i)$, and with local gradient $g(\mathbf{x}_i)$, will not lie in shadow when the surface is illuminated with an incident light ray $\omega_o$ as $S_\mathrm{gen}(f(\mathbf{x}_i), g(\mathbf{x}_i), \omega_o)$, or $S_\mathrm{gen}(\mathbf{x}_i, \omega_o)$ for short, as shown in Figure~\ref{fig:generalized_dv}. It can be written as the limit
\begin{equation}
\label{eqn:shadow_masking}
S_\mathrm{gen}(\mathbf{x}_i, \omega_o)=\lim_{t\to\infty}\mathrm{Tr}_\mathrm{gen}(\mathbf{x}_i, \omega_o, t),
\end{equation}
where $\mathrm{Tr}_\mathrm{gen}(\mathbf{x}_i, \omega_o, t)$ is the probability that no part of the GPIS between $\mathbf{x}_i$ and $\mathbf{x}_i+t\omega_o$ will intersect the ray $\omega_o$. Then $\mathrm{Tr}_\mathrm{gen}(\mathbf{x}_i, \omega_o, t)$ is the transmittance of the ray $\omega_o$ originating from $\mathbf{x}_i$ and advancing $t$. A differential equation of $\mathrm{Tr}_\mathrm{gen}(\mathbf{x}_i, \omega_o, t)$ can be written as:
\begin{equation}
\label{eqn:sq}
\mathrm{Tr}_\mathrm{gen}(\mathbf{x}_i,\omega_o,t+\Delta t)=\mathrm{Tr}_\mathrm{gen}(\mathbf{x}_i,\omega_o,t)W(\Delta t\mid \mathbf{x}_i,\omega_o,t),
\end{equation}
where $W(\Delta t\mid \mathbf{x}_i,\omega_o,t)$ is the conditional probability that the surface will not occlude $\mathbf{x}_i$ in the interval $\Delta t$, given that it does not in the interval $t$. We rewrite it as follows:
\begin{equation}
W(\Delta t\mid \mathbf{x}_i,\omega_o,t)=1-\sigma_t(\mathbf{x}_i, \omega_o, t)\Delta t,
\end{equation}
where $\sigma_t(\mathbf{x}_i, \omega_o, t)\Delta t$ is the conditional probability that the surface in $\Delta t$ will occlude $\mathbf{x}_i$ given that it does not in the interval $t$. We insert it into Equation~\eqref{eqn:sq} and get a differential equation:
\begin{equation}
\frac{\mathrm{d}\mathrm{Tr}_\mathrm{gen}(\mathbf{x}_i,\omega_o,t)}{\mathrm{d}t}=-\sigma_t(\mathbf{x}_i, \omega_o, t)\mathrm{Tr}_\mathrm{gen}(\mathbf{x}_i,\omega_o,t).
\end{equation}
It can be integrated to yield
\begin{equation}
\label{eqn:sm2tr}
\mathrm{Tr}_\mathrm{gen}(\mathbf{x}_i,\omega_o,t)=\exp\left(-\int_0^t\sigma_t(\mathbf{x}_i, \omega_o, t)\mathrm{d}t\right).
\end{equation}
It is worth noting that $\sigma_t(\mathbf{x}_i, \omega_o, t)$ is the extinction coefficient. It depends on the initial point $\mathbf{x}_i$, making the medium non-classic.

We denote circumstance $\alpha$ which is that the surface at $t$ does not occlude $\mathbf{x}_i$, that is,
\begin{equation}
f(\mathbf{x}_i+t\omega_o)>0.
\end{equation}
And we denote circumstance $\beta$ which is that the surface in $\Delta t$ does occlude $\mathbf{x}_i$, that is,
\begin{equation}
\begin{split}
f(\mathbf{x}_i+t\omega_o)&>0,\\
f(\mathbf{x}_i+(t+\Delta t)\omega_o)&<0,\\
k&<0,\\
k&=\omega_o^Tg(\mathbf{x}_i+t\omega_o).
\end{split}
\end{equation}
According to the first-order Taylor series expansion, we can obtain
\begin{equation}
f(\mathbf{x_i}+t\omega_o)<-k\Delta t.
\end{equation}
Then, we can use $\alpha$ and $\beta$ to rewrite the integration in Equation~\eqref{eqn:sm2tr}:
\begin{equation}
\sigma_t(\mathbf{x}_i, \omega_o, t)\Delta t=P(\beta\mid\alpha)=\frac{P(\alpha,\beta)}{P(\alpha)}
\end{equation}
Let $P(f, g\mid \mathbf{x}_i, \omega_o,t)$ be the joint probability density function of $f$ and $g$ at point $\mathbf{x}_{i+1}=\mathbf{x}_i+t\omega_o$, conditional upon given SDF and gradient at $\mathbf{x}_i$, then
\begin{equation}
P(\alpha)=\int_{\mathbb{R}^3}\mathrm{d}g\int_0^{\infty}P(f,g\mid \mathbf{x}_i, \omega_o, t)\mathrm{d}f,
\end{equation}
and
\begin{equation}
\begin{split}
P(\alpha,\beta)&=\int_{-\infty}^0\mathrm{d}k\int_{0}^{-k\Delta t}P(f,g\mid \mathbf{x}_i, \omega_o,t)\mathrm{d}f\\
&=-\Delta t\int_{-\infty}^0kP(0,g\mid \mathbf{x}_i, \omega_o,t)\mathrm{d}k.
\end{split}
\end{equation}
Therefore,
\begin{equation}
\label{eqn:complex_sigma_t}
\sigma_t(\mathbf{x}_i, \omega_o, t)=-\frac{\int_{-\infty}^0kP(0,g\mid \mathbf{x}_i, \omega_o,t)\mathrm{d}k}{\int_{\mathbb{R}^3}\mathrm{d}g\int_0^{\infty}P(f,g\mid \mathbf{x}_i, \omega_o,t)\mathrm{d}f}
\end{equation}

\begin{figure}[t]
    \centering
    \includegraphics[width=0.243\linewidth]{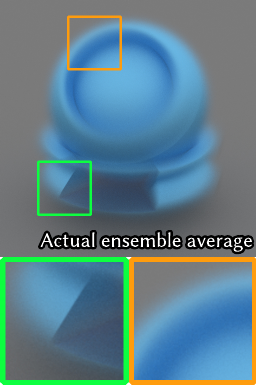}
    \includegraphics[width=0.243\linewidth]{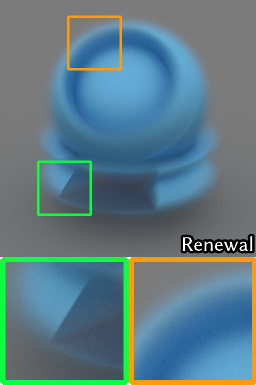}
    \includegraphics[width=0.243\linewidth]{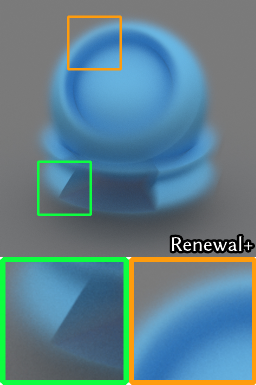}
    \includegraphics[width=0.243\linewidth]{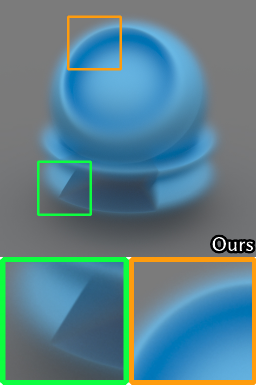}
    \caption{Impact of different independence assumptions. The macrofacet has a similar appearance to the actual ensemble average although there are slight differences (orange insets) at regions of high curvature due to our independence assumption, as discussed in Section~\ref{sec:discuss}.}
    \label{fig:impact_independent}
\end{figure}

Similar to the Smith's assumption \cite{Smith}, we suppose that the SDF $f(\mathrm{x}_i)$ and gradient $g(\mathrm{x}_i)$ at $\mathbf{x}_i$ and those at $\mathbf{x}_{i+1}$ are independent. This is a 3D extension of the Smith's assumption \cite{Smith}, which is widely used by microfacet theory \cite{16Heitz, Position-free, Invariance-principle, 22Bitterli}. Previous GPIS works \cite{GPIS, SCNoise} also assume independence to some extent, as shown in Figure~\ref{fig:indpendent_assumption}, to increase the speed of finding ray intersections. Without this assumption, GPSSes cannot be represented as classic exponential participating media essentially, as discussed in Section~\ref{sec:discuss}. The impact of this assumption is shown in Figure~\ref{fig:impact_independent}.

After we apply this independence assumption and Equation~\eqref{eqn:joint_fg}, the joint probability becomes
\begin{equation}
\begin{split}
P(f,g\mid \mathbf{x}_i, \omega_o,t)=&P(f, g)\\
=&\phi(f; 0, \sigma^2)\phi\left(g; (0, 0, 1)^\top, \sigma^2\mathrm{diag}(l_x^2, l_y^2, l_z^2)^{-1}\right).
\end{split}
\end{equation}
Note that the SDF $f$ and gradient $g$ are naturally independent as the result of Equation~\eqref{eqn:joint_fg}. We insert this result into Equation~\eqref{eqn:complex_sigma_t} and get
\begin{equation}
\begin{split}
\label{eqn:generalized_sigma_t}
\sigma_t(\omega_o, t)=&\rho(t)\sigma(\omega_o),\\
\rho(t)=&\frac{\phi(z_i+t\cos\theta_o; 0, \sigma^2)}{\Phi(z_i+t\cos\theta_o; 0, \sigma^2)},\\
\sigma(\omega_o)=&\Lambda(\omega_o)\cos\theta_o,
\end{split}
\end{equation}
\begin{equation}
\begin{split}
\label{eqn:generalized_lambda}
\Lambda(\omega_o)=&\frac{1}{2a\sqrt{\pi}}e^{-a^2}+\frac{1}{2}(\mathrm{erf}(a)-1),\\
a=&\frac{\cos\theta_o}{\sqrt{\alpha_x^2\sin^2\theta_o\cos^2\phi_o+\alpha_y^2\sin^2\theta_o\sin^2\phi_o+\alpha_z^2\cos^2\theta_o}},
\end{split}
\end{equation}
where $\mathrm{erf}$ is the error function, and $z_i+t\cos\theta_o$ is in fact the $z$ coordinate $z_{i+1}$ of the next intersection. Here we define $\alpha_z=\sqrt{2}\sigma/l_z$ as the ``roughness'' on the $z$-axis. When $\alpha_z\to 0$, that is, $l_z\to\infty$, the GPSS is a height field, and Equation~\eqref{eqn:generalized_lambda} degenerates to the Beckmann version of the Smith's Lambda function. The extinction coefficient no longer depends on the initial point because of our independence assumption. Notably, the form of the extinction coefficient is the same as Equation~\eqref{eqn:macro_sigma_t}, where $\rho(t)$ is the density and $\sigma(\omega_o)$ is the projected area, maintaining the consistency. Figure~\ref{fig:tr} compares the transmittance of our generalized macrofacet and an isotropic GPSS. When the travel distance $t$ increases, there will be a slight difference from the transmittance of the GPSS. We believe that it is because of the de-correlation to simply calculate the extinction coefficient. However, if we do not assume the independence, the extinction coefficient will depend on the origin of the ray, making the macrofacet not a classic exponential participating medium.

\subsubsection{Phase function}
Because the GPSS is not a height field any more, some equivalences in microfacet theory are no longer applicable, such as the preservation of the projected area of microfacets \cite{Heitz2014Microfacet}:
\begin{equation}
\int_{H^2} \langle \omega_m, \omega_g\rangle D(\omega_m)\mathrm{d}\omega_m=1.
\end{equation}
When the GPSS is not a height field, its normals are distributed on the full sphere instead of the hemisphere. Also, there are normals pointing downwards because of holes and overlaps in the micro-geometry, affecting the NDF and making the integration on the left hand side not equal to 1. In the following, we derive an analytical solution of this general NDF using the gradient distribution so that we can use it to compute the phase function.

\begin{figure}[t]
    \centering
    \includegraphics[width=0.9\linewidth]{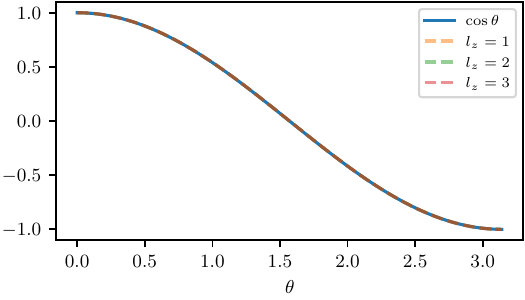}
    \caption{The equivalence between the signed projected area of micro-surface and macro-surface. We compare the signed projected area of the micro-surface computed by numerical integration (yellow, green and red lines) and the one of the macro-surface (blue line) in Equation~\eqref{eqn:spa} with different correlation along the $z$-axis $l_z$. The signed projected area of the micro-surface matches the one of the macro-surface well regardless of the value of $l_z$.}
    \label{fig:spa}
\end{figure}

\begin{figure}[t]
    \centering
    \includegraphics[width=0.9\linewidth]{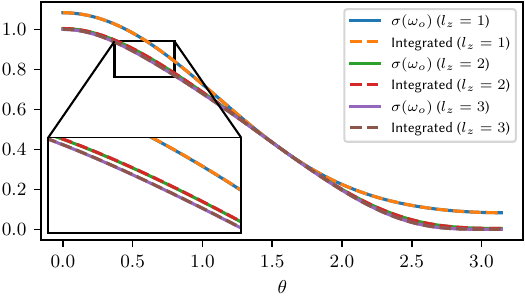}
    \caption{The equivalence between the normalized factor in vNDF and projected area. We compare the normalized factor computed by numerical integration (orange, red and brown lines) and the projected area $\sigma(\omega_o)$ in Equation~\eqref{eqn:generalized_sigma_t} (blue, green and purple lines) with different correlation along the $z$-axis $l_z$. The normalized factor matches projected area well for all values of $l_z$.}
    \label{fig:generalized_projected_area}
\end{figure}

\begin{figure*}[t]
    \centering
    \includegraphics[width=0.96\linewidth]{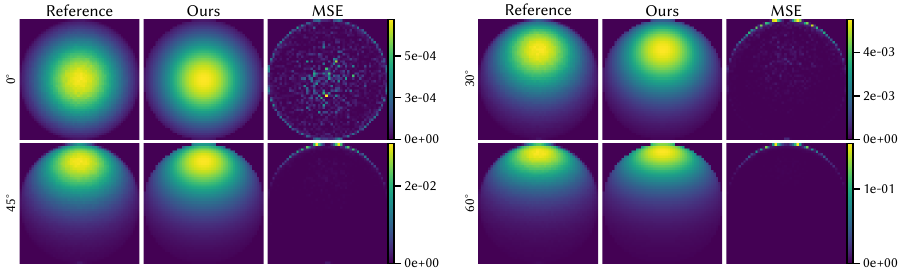}
    \caption{We compare our vNDFs with the one generated by GPISes at different incident angles. The references are vNDFs generated by GPISes. The roughness on all axes is $1.0$. The MSE shows that our vNDFs match references accurately.}
    \label{fig:vndf}
\end{figure*}

Since the normal $\omega_m$ is the normalized gradient $g$, we can connect them with the magnitude $t$ of $g$. Hence, the NDF is the marginal of the gradient distribution function (GDF) $D_g$ over all gradient magnitudes:
\begin{equation}
\label{eqn:marginal_gdf}
D(\omega_m)=\int_0^\infty D_g(g)t^2\mathrm{d}t,
\end{equation}
where $g=t\omega_m$, and $t^2$ is the Jacobian factor. Notably, we use the term GDF to distinguish it from the probability density function (PDF) $P(g)$ of $g$ as they are not the same. Recall the definition of the NDF \cite{GGX}: the NDF multiplied by an infinitesimal solid angle $\mathrm{d}\omega_m$ centered on $\omega_m$ and an infinitesimal macro-surface area $\mathrm{d}A$ is the total area of the portion of the corresponding micro-surface whose normals lie within $\mathrm{d}\omega_m$. Therefore, the NDF is the area of micro-surfaces. Similarly, the GDF is the area-weighted distribution of gradients:
\begin{equation}
\label{eqn:pdf_to_gdf}
D_g(g)=tP(g).
\end{equation}
Here, $t$ also acts as the area of the gradient $g$. We insert it into Equation~\eqref{eqn:marginal_gdf} and get:
\begin{equation}
\label{eqn:gdf_to_ndf}
D(\omega_m)=\int_0^{\infty}P(g)t^3\mathrm{d}t.
\end{equation}
Within this equation, we establish the connection between the PDF of gradients and the NDF. It does not depend on Gaussian processes. It is applicable to any PDFs of gradients, and also SDF distributions.

Next, we need to know about the PDF of gradients. Within the independence assumption we make in the transmittance calculation, we do not need to consider other points but the next intersection $\mathbf{x}_{i+1}$ on the ray. Therefore, the distribution $P(g(\mathrm{x}_{i+1})\mid f(\mathbf{x}_{i+1})=0)$ of the gradient $g(\mathbf{x}_{i+1})$ conditioned on $f(\mathbf{x}_{i+1})=0$ is a Gaussian distribution as the result of Equation~\eqref{eqn:joint_fg} and Equation~\eqref{eqn:condition_gp}:
\begin{equation}
\label{eqn:gdf}
g' \sim \mathcal{N}\left((0, 0, 1)^T, \sigma^2\mathrm{diag}(l_x^2, l_y^2, l_z^2)^{-1}\right),
\end{equation}
where we simplify the notation $g(\mathrm{x}_{i+1})\mid f(\mathbf{x}_{i+1})=0$ as $g'$. We insert it into Equation~\eqref{eqn:gdf_to_ndf}. After the simplification we show in Appendix~\ref{sec:ndf_gen}, we can get
\begin{equation}
\begin{split}
\label{eqn:macro_ndf}
D(\omega_m)=&\frac{e^{-C+B^2/A}}{\pi^{3/2}\alpha_x\alpha_y\alpha_z}\Bigg[\frac{1}{2A^2}\left(\frac{B^2}{A}+1\right)e^{-B^2/A}\\&+\frac{B\sqrt{\pi}}{2A^{5/2}}\left(\frac{B^2}{A}+\frac{3}{2}\right)\mathrm{erfc}\left(-\frac{B}{\sqrt{A}}\right)\Bigg],\\
A=&\frac{\sin^2\theta_m\cos^2\phi_m}{\alpha_x^2}+\frac{\sin^2\theta_m\sin^2\phi_m}{\alpha_y^2}+\frac{\cos^2\theta_m}{\alpha_z^2},\\
B=&\frac{\cos\theta_m}{\alpha_z^2},\\
C=&\frac{1}{\alpha_z^2},
\end{split}
\end{equation}
where $\mathrm{erfc}$ is the complementary error function. When $\alpha_z\to 0$, that is, $l_z\to \infty$, Equation~\eqref{eqn:macro_ndf} degenerates to the Beckmann NDF. Interestingly, some equivalences about NDF in microfacet theory still hold. For example, the signed projected area of the micro-surface is the same as the projected area of the macro-surface for any direction $\omega$:
\begin{equation}
\label{eqn:spa}
\int_{\Omega} (\omega\cdot\omega_m)D(\omega_m)\mathrm{d}\omega_m=(\omega\cdot\omega_m).
\end{equation}
We show this equivalence in Figure~\ref{fig:spa}. This ensures that the light cannot pass through the surface.

As we have the analytical solution for the NDF, we can compute the vNDF using Equation~\eqref{eqn:microfacet_vndf} so as to compute the phase function using Equation~\eqref{eqn:phase}. However, the normalized factor in the denominator is still a problem. In fact, it is equal to the projected area listed in Equation~\eqref{eqn:generalized_sigma_t}. We show this equivalence in Figure~\ref{fig:generalized_projected_area}. This maintains the consistency with microflake theory and the reciprocity of the phase function. We compare our vNDF with the one generated by the GPIS \cite{GPIS} in Figure~\ref{fig:vndf}, showing the great consistency.

\begin{figure}[t]
        \centering
        \includegraphics[width=0.75\linewidth]{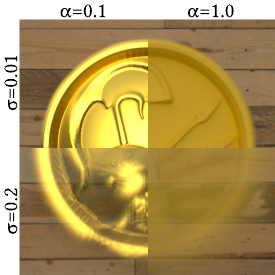}
        \caption{Macrofacet with different variance and correlation.}
        \label{fig:spa_macro}
    \end{figure}
    \begin{figure}[t]
        \centering
        \begin{minipage}{0.49\linewidth}
            \centering
            $\alpha_z = 0.01$
            \includegraphics[width=\linewidth]{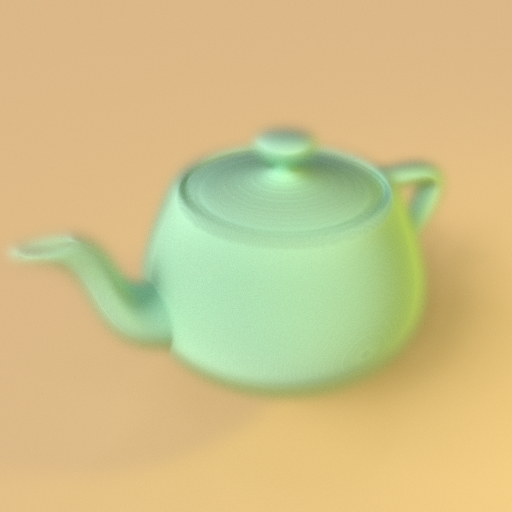}
        \end{minipage}
        \begin{minipage}{0.49\linewidth}
            \centering
            $\alpha_z = 0.8$ 
            \includegraphics[width=\linewidth]{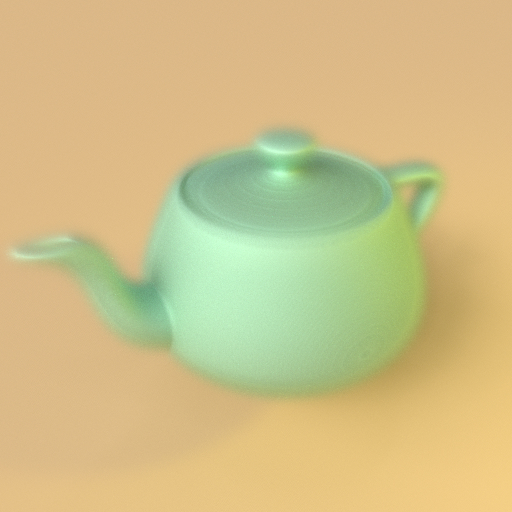}
        \end{minipage}
        
        \caption{Comparisons between the anisotropic generalized macrofacet (left) and the isotropic generalized macrofacet (right). The highlight and fuzziness show the impact of the correlation along the $z$-axis. This is a case that microfacet theory \cite{16Heitz, Dupuy} cannot handle.}
        \label{fig:cmp_aniso}
    \end{figure}

Importance sampling this vNDF is a difficult problem due to the error function making it unable to use inverse methods. For now, we combine the Beckmann vNDF importance sampling \cite{vndf-is} with ratio $R$ and uniform hemisphere whose $z$-axis is $-\omega_o$ sampling with ratio $1-R$ to sample $\omega_m$. The PDF of sampling $\omega_m$ is
\begin{equation}
P(\omega_m)=RP_\mathrm{Beckmann}(\omega_m)+(1-R)P_\mathrm{uniform}(\omega_m),
\end{equation}
where $P_\mathrm{Beckmann}$ is the PDF of the Beckmann vNDF importance sampling and $P_\mathrm{uniform}(\omega_m)$ is the uniform hemisphere sampling.

\begin{figure*}[t]
    \centering
    \includegraphics[width=0.9\linewidth]{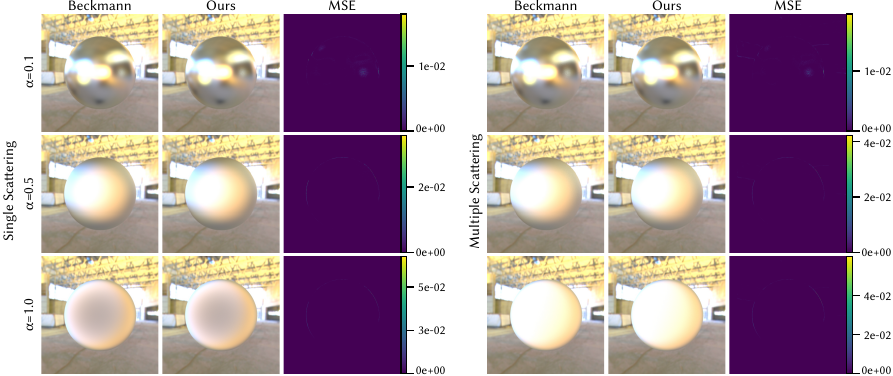}
    \caption{Comparisons between Beckmann macrofacets and Beckmann microfacets with different roughness. We conduct comparisons for single scattering and multiple scattering, respectively. The MSE shows that the macrofacet is consistent with the microfacet.}
    \label{fig:cmp_microfacet}
\end{figure*}

\begin{figure*}[t]
    \centering
    \includegraphics[width=0.83\linewidth]{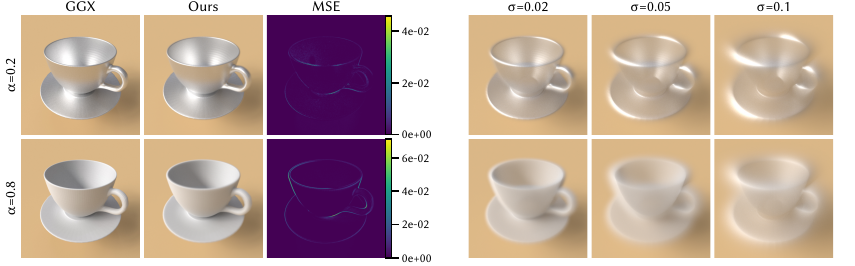}
    \caption{Comparisons between GGX macrofacets and GGX microfacets with different roughness. Columns four through six correspond to cases with varying variance. The MSE shows that the macrofacet can adapt to different NDFs.}
    \label{fig:cmp_ggx}
\end{figure*}

\begin{figure*}[t]
    \centering
    \includegraphics[width=0.9\linewidth]{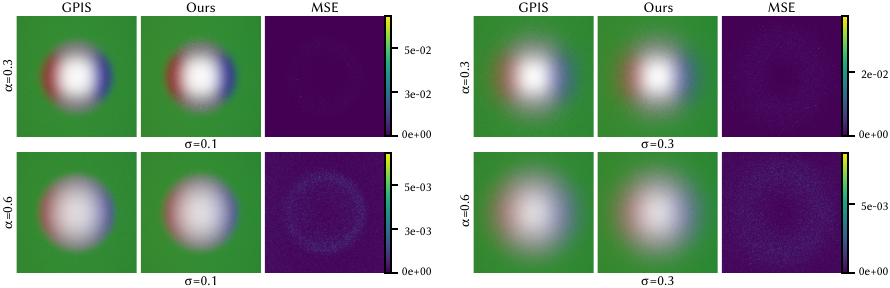}
    \caption{Comparisons between macrofacets and GPISes with different pairs of variance and roughness. The MSE shows that the macrofacet preserves statistical consistency with the GPIS.}
    \label{fig:cmp_gp}
\end{figure*}

\section{RESULTS}

We implemented our macrofacet in PBRT \cite{PBRTV3}. We ran all experiments with an Intel Core i9-9900K CPU. We adopted null-scattering \cite{null-scattering} as our volumetric rendering method\footnote{In fact, any classic exponential participating media rendering method can be chosen.}. We used the mean squared error (MSE) to evaluate the difference between rendering results and references. Our results show high quality renderings at a low time cost. \emph{Note that all references in Section~\ref{sec:cmp_gp} are the converged results by Xu et al.~\shortcite{SCNoise}, showing that our results are very similar to theirs.} We show a macrofacet with different parameter settings in Figure~\ref{fig:spa_macro}.

\subsection{Comparisons with Microfacet}

\begin{figure*}[t]
    \centering
    \includegraphics[width=0.24\linewidth]{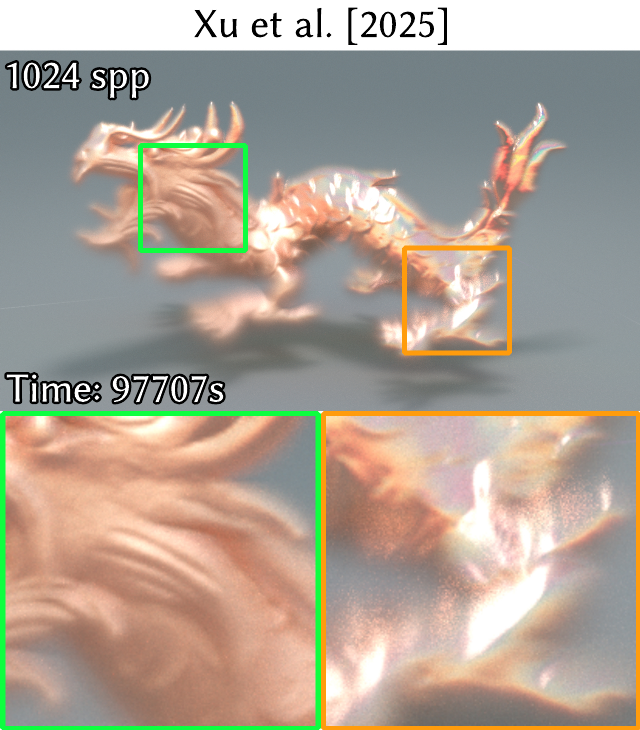}
    \includegraphics[width=0.24\linewidth]{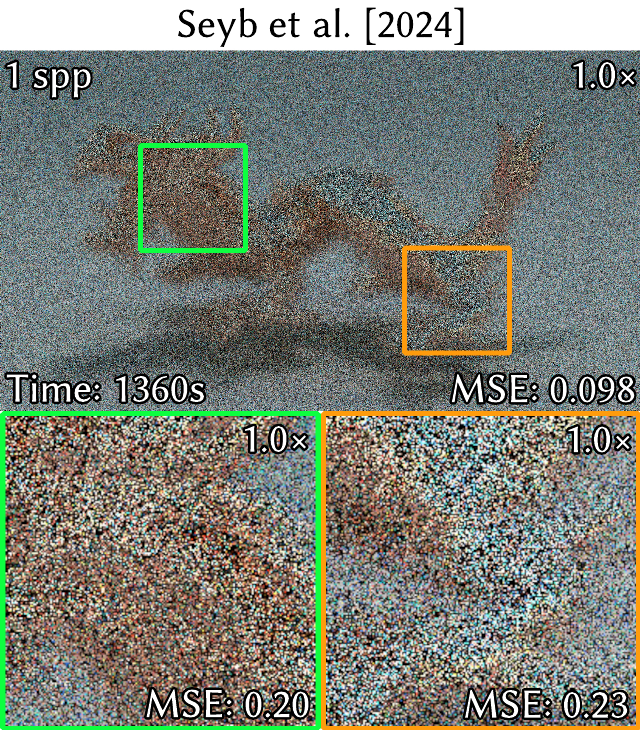}
    \includegraphics[width=0.24\linewidth]{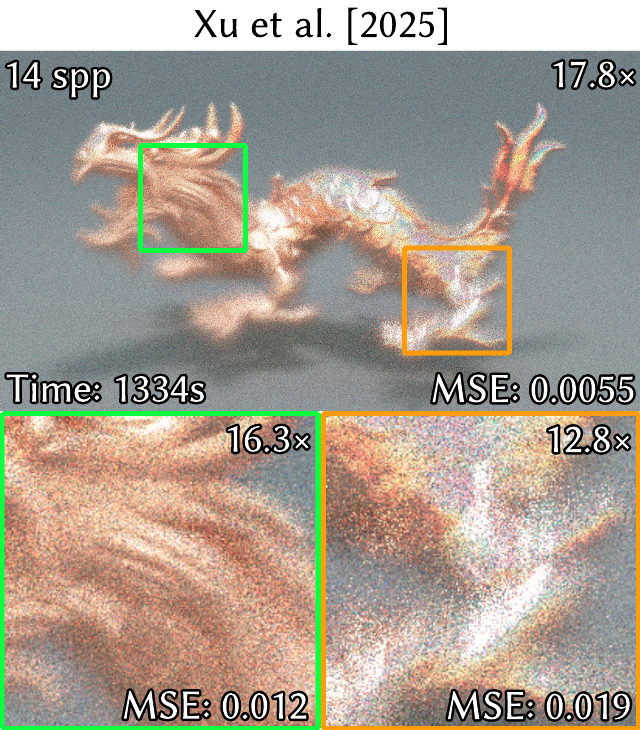}
    \includegraphics[width=0.24\linewidth]{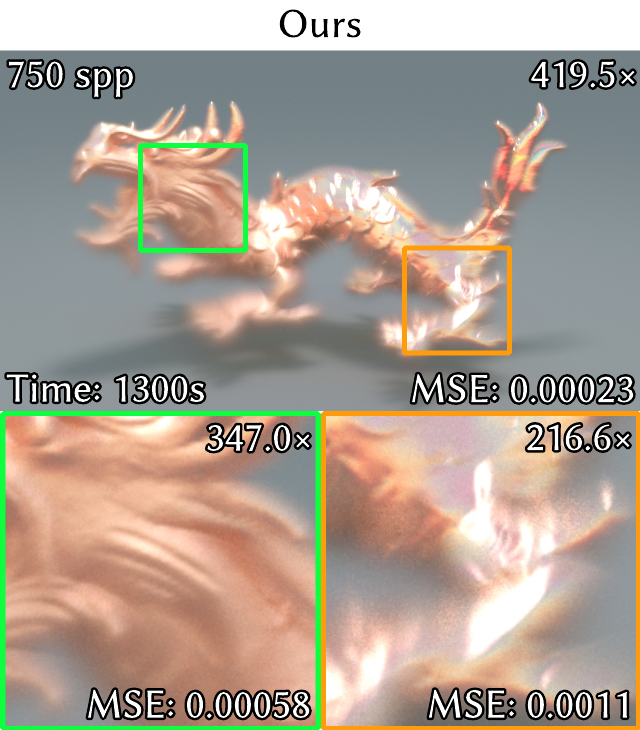}
    \caption{We evaluate previous GPIS approaches \cite{GPIS, SCNoise} and the macrofacet at equal time (about 22 min). The variance of the non-stationary kernel increases from top to bottom, while the correlation of it increases from left to right. The leftmost column shows a reference rendered by Xu et al.~\shortcite{SCNoise}. The spp, time, MSE and speedup for the full image and for image patches are shown in corners.}
    \label{fig:cmp_dragon}
\end{figure*}

\begin{figure*}[t]
    \centering
    \includegraphics[width=0.24\linewidth]{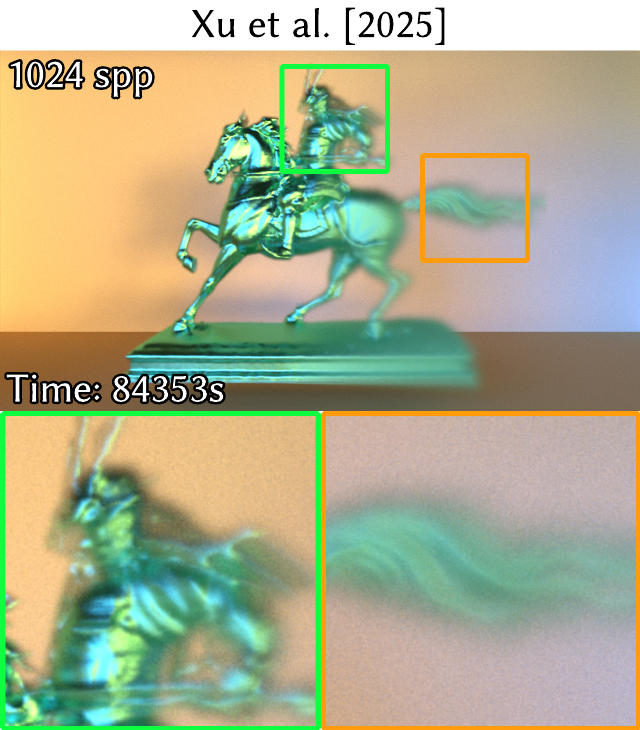}
    \includegraphics[width=0.24\linewidth]{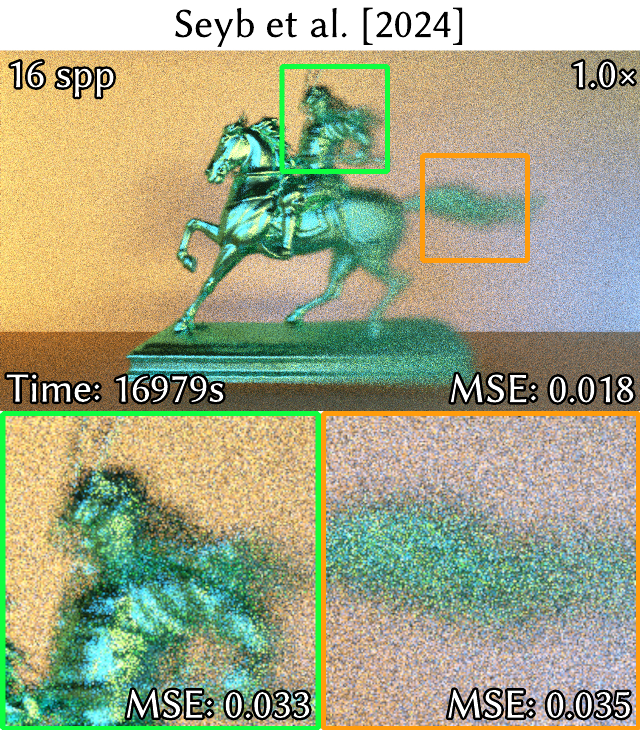}
    \includegraphics[width=0.24\linewidth]{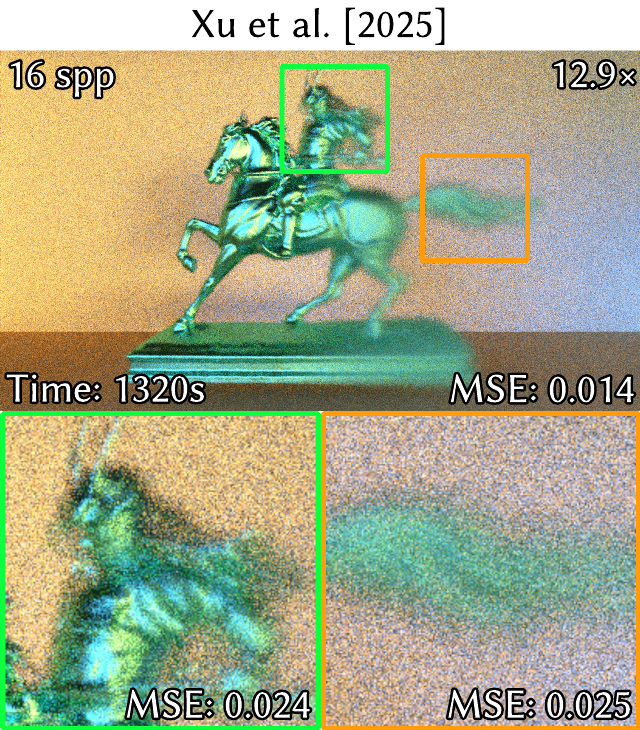}
    \includegraphics[width=0.24\linewidth]{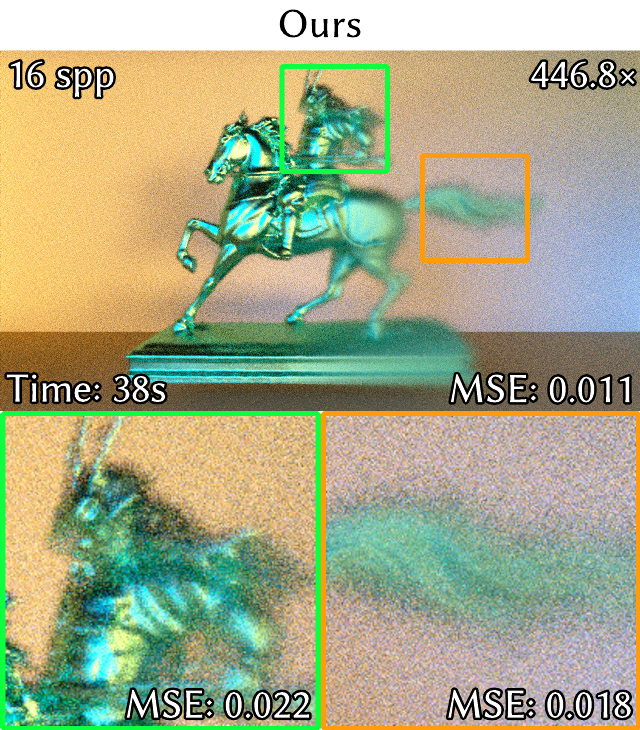}
    \caption{We evaluate previous GPIS approaches \cite{GPIS, SCNoise} and the macrofacet at equal samples per pixel (16 spp). The variance of the non-stationary kernel increases from left to right, while the correlation of it increases from bottom to top. The leftmost column shows a reference rendered by Xu et al.~\shortcite{SCNoise}. The spp, time, MSE and speedup for the full image and for image patches are shown in corners.}
    \label{fig:cmp_horse}
\end{figure*}

\begin{figure}[t]
    \centering
    \includegraphics[width=0.7\linewidth]{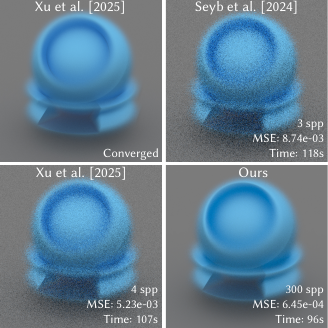}
    \caption{Comparisons between the macrofacet and the GPIS at equal time. It shows that the macrofacet converges much faster than others.}
    \label{fig:cmp_et}
\end{figure}

We compare the Beckmann macrofacet with the Beckmann microfacet in Figure~\ref{fig:cmp_microfacet}. We set the variance $\sigma=0.01$ to limit the length of the shell to be like a surface. The roughness changes from 0.1 to 1.0. We render results in both single scattering and multiple scattering. The multiple scattering Beckmann microfacet is implemented from Cui et al.~\shortcite{Invariance-principle}. The results show that the macrofacet is consistent with the microfacet in both single scattering and multiple scattering. It also means that the macrofacet supports the multiple scattering microfacet naturally because it is rendered as a volume. We also compare the GGX macrofacet with the GGX microfacet in Figure~\ref{fig:cmp_ggx}. It shows that no matter what NDF is chosen, our macrofacet fits with the microfacet when it degenerates to a surface. There are minor differences on edges of the model because the length of the shell cannot be omitted. Besides, Figure~\ref{fig:cmp_ggx} shows how the appearance of the macrofacet changes when the variance $\sigma$ increases. We proceed to compare the anisotropic generalized macrofacet and the isotropic generalized macrofacet in Figure~\ref{fig:cmp_aniso}, showing the impact of correlation along the $z$-axis.

\subsection{Comparisons with Gaussian Processes}
\label{sec:cmp_gp}

We compare the generalized macrofacet with isotropic GPISes proposed by Seyb et al.~\shortcite{GPIS} and Xu et al.~\shortcite{SCNoise} with different variance and roughness. All GPISes use the squared exponential kernel as their covariance kernel. We keep the variance and roughness of GPISes the same as those of the macrofacet. We first verify whether macrofacet appears similar to the GPIS with different pairs of variance and roughness. Figure~\ref{fig:cmp_gp} shows that no matter what variance and roughness it is, our macrofacet appears similar to GPIS.

We render generalized macrofacet and previous GPIS approaches \cite{GPIS, SCNoise} at equal time (100 seconds) in Figure~\ref{fig:cmp_et}. The result shows that the macrofacet is significantly faster than GPIS approaches and can render more samples per pixel (spp) at the same time because it is realization free.

Moreover, we compare the generalized macrofacet with spatially variant GPISes in Figure~\ref{fig:cmp_dragon} and Figure~\ref{fig:cmp_horse}. We use the converged results rendered by Xu et al.~\shortcite{SCNoise} as references. At equal time, the macrofacet has already converged when Seyb et al.~\shortcite{GPIS} only render 1 spp. Therefore, the MSE can also be viewed as the difference due to our independence assumption. At equal samples per pixel, the macrofacet achieves an order-of-magnitude improvement in rendering speed. Furthermore, the macrofacet converges faster than others within the same spp because we trace statistically rather than realize implicit surfaces individually for each path. As a consequence, the macrofacet significantly outperforms previous work in rendering time and maintains a similar appearance.

\begin{figure}[t]
    \centering
    \includegraphics[width=0.9\linewidth]{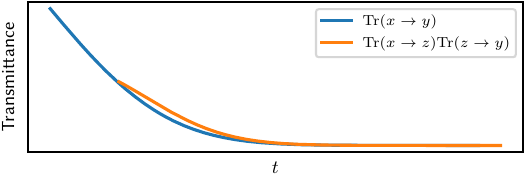}
    \captionof{figure}{We show two lines of transmittance of a GPSS starting from $x$ (blue line) and $z$ (orange line). We then multiply the transmittance started from $z$ by the transmittance $\mathrm{Tr}(x\to z)$ from $x$ to $z$. For any point $y$ after $z$, these two lines do not match, which means that the transmittance is not multiplicative: $\mathrm{Tr}(x\to y)\neq\mathrm{Tr}(x\to z)\mathrm{Tr}(z\to y)$. Therefore, the GPSS is not a classic exponential participating medium.}
    \label{fig:multiplicative_transmittance}
\end{figure}

\section{DISCUSSION}
\label{sec:discuss}

\paragraph{Independence assumptions}
Both the macrofacet and previous GPIS approaches \cite{GPIS, SCNoise} make different relaxations to deal with dependency because of the correlation. Previous GPIS approaches march rays segment by segment. Because a realization of a Gaussian process conditioned on the entire path costs unacceptable time, they propose Renewal and Renewal+ models. The Renewal model only conditions on the SDF of the previous intersection, while the Renewal+ model conditions on the SDF and gradient of the previous intersection. In other words, they both make an assumption that the current realization is independent of earlier segments and intersections. The macrofacet is neither Renewal nor Renewal+ model. We assume that the SDF and gradient at the next intersection are independent of those at the last intersection. The difference is shown in Figure~\ref{fig:indpendent_assumption}. Because of this, we do not expect exactly the same result. We believe that the macrofacet is promising to handle full path correlation if we introduce non-classic media.

\paragraph{Non-classic media}
The differences between the actual ensemble average and ours in Figure~\ref{fig:impact_independent} appear at regions of high curvature. Our independence assumption drops the information that the ray starts at the previous intersection, and this information only decays as the ray travels away from it. When the surface is flat, the ray leaves it quickly and the information has already decayed before the next intersection, so the assumption is nearly exact. When the surface is highly curved, the ray stays close to it and often intersects it again nearby, so the dropped information still matters. GPSSes cannot be represented as classic exponential participating media essentially because of this correlation. The correlation makes the volumetric properties, such as the extinction coefficient, of the current point depend on the origin point of the ray, as discussed in Section~\ref{sec:generalized_macrofacet}. As a result, the transmittance is not multiplicative along points on a ray, where $\mathrm{Tr}(x\to y)\neq\mathrm{Tr}(x\to z)\mathrm{Tr}(z\to y)$, as shown in Figure~\ref{fig:multiplicative_transmittance}. Multiplicative transmittance is an important characteristic of classic exponential participating media. Hence, classic exponential participating media cannot handle such correlation and we make the independence assumption. We believe that non-exponential participating media \cite{bitterli-non-exp, jarabo-media} can be used to deal with such correlation to achieve an appearance similar to the Renewal+ model, because all of them condition on the previous intersection. However, classic exponential participating media have already appeared very similar to the actual ensemble average in practice, as shown in Figure~\ref{fig:impact_independent}. Therefore, we leave non-exponential participating media solutions to future work.

\paragraph{Normal distribution}
Inspired by the definition of the NDF, we propose the GDF in Equation~\eqref{eqn:pdf_to_gdf} to compute the NDF. This is independent of Gaussian processes. Furthermore, within the independence assumption, we can use the SDF distribution to define the micro-geometry and convert any stochastic processes implicit surface to the macrofacet if we know the GDF and SDF distribution of any point in the space.

\section{CONCLUSION AND FUTURE WORK}

We present macrofacet theory. We extend a macro-surface to a volume according to the variance and convert microfacet theory into microflake theory in the macro-space. Within macrofacet theory, microfacet theory is connected to the GPSS. In order to support correlation along the $z$-axis, which we call the generalized macrofacet, we apply an independence assumption to analytically compute the extinction coefficient and phase function. This independence assumption, as shown in Figure~\ref{fig:indpendent_assumption}, is the only assumption we make to represent GPSSes as classic exponential participating media. As a consequence, we can render the appearance of the GPSS by classic exponential participating media rendering methods to accomplish a faster convergence speed. Moreover, artists can choose different typical normal distributions to achieve the appearance they want without understanding principles of Gaussian processes.

In the future, we would use non-exponential participating media to represent GPSSes to solve correlation problems like non-multiplicative transmittance and full spherical visible normals' distribution. Besides, since noise is produced from volumetric rendering, which has extensive research compared to GPISes, we would implement denoising techniques to further improve our convergence in order to achieve real-time rendering. In addition, we would try to use importance sampling to sample the vNDF of the generalized macrofacet. Moreover, we only focus on the conductor material in this paper, so we would investigate the dielectric material later.

\section{ACKNOWLEDGEMENTS}

We would like to thank all anonymous reviewers for their valuable feedback and guidance. The models in Figure~\ref{fig:teaser} are obtained from \href{https://www.3d66.com/}{3d66}. The knob scenes (Figures~\ref{fig:impact_independent} and \ref{fig:cmp_et}) and the dragon scene (Figure~\ref{fig:cmp_dragon}) are adapted from the scenes in Seyb et al.~\shortcite{GPIS}. Figure~\ref{fig:spa_macro} uses a model from \href{https://www.cgtrader.com/}{CGTrader}. The Utah Teapot in Figure~\ref{fig:cmp_aniso} is provided by the \href{https://graphics.cs.utah.edu/}{Utah Graphics Lab}. The models in Figures~\ref{fig:cmp_ggx} and \ref{fig:cmp_horse} are sourced from \href{https://sketchfab.com/}{Sketchfab}.


\bibliographystyle{ACM-Reference-Format}
\bibliography{macrofacet-bibliography}

\appendix
\section{NORMAL DISTRIBUTION FUNCTION OF GENERALIZED MACROFACET}
\label{sec:ndf_gen}
In this section, we show the simplification of Equation~\eqref{eqn:gdf_to_ndf}.
\begin{equation}
\begin{split}
D(\omega_m)=&\int_0^{\infty}P(g)t^3\mathrm{d}t\\
=&\frac{1}{\pi^{3/2}\alpha_x\alpha_y\alpha_z}\int_0^{\infty}t^3\exp\Bigg(-\frac{t^2\sin^2\theta_m\cos^2\phi_m}{\alpha_x^2}\\&-\frac{t^2\sin^2\theta_m\sin^2\phi_m}{\alpha_y^2}-\frac{(t\cos\theta_m-1)^2}{\alpha_z^2}\Bigg)\mathrm{d}t\\
=&\frac{1}{\pi^{3/2}\alpha_x\alpha_y\alpha_z}\int_0^{\infty}t^3\exp\Bigg(-\Bigg(\frac{\sin^2\theta_m\cos^2\phi_m}{\alpha_x^2}\\&+\frac{\sin^2\theta_m\sin^2\phi_m}{\alpha_y^2}+\frac{\cos^2\theta_m}{\alpha_z^2}\Bigg)t^2+\frac{2\cos\theta_m}{\alpha_z^2}t-\frac{1}{\alpha_z^2}\Bigg)\mathrm{d}t.
\end{split}
\end{equation}
We denote
\begin{equation}
\begin{split}
A=&\frac{\sin^2\theta_m\cos^2\phi_m}{\alpha_x^2}+\frac{\sin^2\theta_m\sin^2\phi_m}{\alpha_y^2}+\frac{\cos^2\theta_m}{\alpha_z^2},\\
B=&\frac{\cos\theta_m}{\alpha_z^2},\\
C=&\frac{1}{\alpha_z^2}.
\end{split}
\end{equation}
Then,
\begin{equation}
\begin{split}
\label{eqn:d_t}
D(\omega_m)=&\frac{e^{-C}}{\pi^{3/2}\alpha_x\alpha_y\alpha_z}\int_0^{\infty}t^3\exp(-At^2+2Bt)\mathrm{d}t\\
=&\frac{e^{-C+B^2/A}}{\pi^{3/2}\alpha_x\alpha_y\alpha_z}\int_0^{\infty}t^3\exp\left(-A\left(t-\frac{B}{A}\right)^2\right)\mathrm{d}t.
\end{split}
\end{equation}
Let $u=t-B/A$. Then $t=u+B/A$ and $\mathrm{d}t=\mathrm{d}u$. We can rewrite Equation~\eqref{eqn:d_t} as:
\begin{equation}
\begin{split}
\label{eqn:d_u}
D(\omega_m)=&\frac{e^{-C+B^2/A}}{\pi^{3/2}\alpha_x\alpha_y\alpha_z}\int_{-B/A}^{\infty}\left(u+\frac{B}{A}\right)^3e^{-Au^2}\mathrm{d}u.
\end{split}
\end{equation}
We denote the integration as $I$:
\begin{equation}
\begin{split}
I=&\int_{-B/A}^\infty \left(u^3+\frac{3B}{A}u^2+\frac{3B^2}{A^2}u+\frac{B^3}{A^3}\right)e^{-Au^2}\mathrm{d}u\\
=&I_1+I_2+I_3+I_4,\\
I_1=&\int_{-B/A}^\infty u^3e^{-Au^2}\mathrm{d}u,\\
I_2=&\frac{3B}{A}\int_{-B/A}^\infty u^2e^{-Au^2}\mathrm{d}u,\\
I_3=&\frac{3B^2}{A^2}\int_{-B/A}^\infty ue^{-Au^2}\mathrm{d}u,\\
I_4=&\frac{B^3}{A^3}\int_{-B/A}^\infty e^{-Au^2}\mathrm{d}u.
\end{split}
\end{equation}
As we simplify $I_1$, $I_2$, $I_3$ and $I_4$ individually, we obtain:
\begin{equation}
\begin{split}
I_1=&\left(\frac{B^2}{2A^3}+\frac{1}{2A^2}\right)e^{-B^2/A},\\
I_2=&-\frac{3B^2}{2A^3}e^{-B^2/A}+\frac{3B\sqrt{\pi}}{4A^{5/2}}\mathrm{erfc}\left(-\frac{B}{\sqrt{A}}\right),\\
I_3=&\frac{3B^2}{2A^3}e^{-B^2/A},\\
I_4=&\frac{B^3\sqrt{\pi}}{2A^{7/2}}\mathrm{erfc}\left(-\frac{B}{\sqrt{A}}\right).
\end{split}
\end{equation}
We add up $I_1$, $I_2$, $I_3$ and $I_4$ to get $I$ and insert it into Equation~\eqref{eqn:d_u}. Then we obtain the final result:
\begin{equation}
\begin{split}
D(\omega_m)=&\frac{e^{-C+B^2/A}}{\pi^{3/2}\alpha_x\alpha_y\alpha_z}\Bigg[\frac{1}{2A^2}\left(\frac{B^2}{A}+1\right)e^{-B^2/A}\\&+\frac{B\sqrt{\pi}}{2A^{5/2}}\left(\frac{B^2}{A}+\frac{3}{2}\right)\mathrm{erfc}\left(-\frac{B}{\sqrt{A}}\right)\Bigg],
\end{split}
\end{equation}
which is the same as Equation~\eqref{eqn:macro_ndf}.

\end{document}